\documentclass{article}

\usepackage{arxiv}

\usepackage[utf8]{inputenc} 
\usepackage[T1]{fontenc}    
\usepackage{hyperref}       
\usepackage{url}            
\usepackage{amsfonts}       
\usepackage{nicefrac}       
\usepackage{microtype}      
\usepackage{lipsum}
\usepackage{graphicx}
\graphicspath{ {./images/} }
\usepackage{newunicodechar}
 \newunicodechar{≥}{\geq}

\usepackage{graphicx}
\usepackage{color}
\usepackage{tabularx}
\usepackage{enumitem}
\usepackage{algorithm}
\usepackage{algorithmic}
\usepackage{amsmath, amssymb}
\usepackage{subcaption}
\usepackage{hyperref}
\usepackage{cleveref}
\usepackage{comment}
\usepackage[table,xcdraw]{xcolor}  
\usepackage{colortbl}              

\title{JSPIM: A Skew-Aware PIM Accelerator for High-Performance Databases Join and Select Operations}

\author{
 Sabiha Tajdari \\
 Computer Science Department\\
  University of Virginia\\
  \texttt{jvx2tt@virginia.edu} \\
   \And
 Anastasia Ailamaki \\
  School of Computer and Communication Sciences\\
  École Polytechnique Fédérale de Lausanne\\
  \texttt{anastasia.ailamaki@epfl.ch} \\
  \And
  Sandhya Dwarkadas \\
  Computer Science Department\\
  University of Virginia \\
  \texttt{sandhya@virginia.edu} \\
}

\begin{document}
\maketitle
\begin{abstract}
Database applications are increasingly bottlenecked by memory bandwidth and latency due to the memory wall and the limited scalability of DRAM. Join queries, central to analytical workloads, require intensive memory access and are particularly vulnerable to inefficiencies in data movement. While Processing-in-Memory (PIM) offers a promising solution, existing designs typically reuse CPU-oriented join algorithms, limiting parallelism and incurring costly inter-chip communication. Additionally, data skew—a main challenge in CPU-based joins—remains unresolved in current PIM architectures.

We introduce \textbf{JSPIM}, a PIM module that accelerates hash join and, by extension, corresponding select queries through algorithm–hardware co‐design. JSPIM deploys parallel search engines within each subarray and redesigns hash tables to achieve $O(1)$ lookups, fully exploiting PIM’s fine‐grained parallelism. To mitigate skew, our design integrates subarray‐level parallelism with rank‐level processing, eliminating redundant off‐chip transfers.
Evaluations show JSPIM delivers \textbf{400×–1000×} speedup on join queries versus DuckDB. When paired with DuckDB for the full SSB benchmark, JSPIM achieves an overall \textbf{2.5×} throughput improvement (individual query gains of 1.1×–28×), at just a 7\% data overhead and 2.1\% per‐rank PIM-enabed chip area increase.
\end{abstract}

\section{{Introduction}}
The widening gap between CPU and memory performance, commonly known as the memory wall~\cite{memwall}, has a significant impact on data-intensive applications, which are becoming more prevalent.
This gap is exacerbated by the challenges of DRAM bandwidth scaling \cite{memcentricDB}, increasing the cost of data transfer between the CPU and DRAM memory. 
To address these challenges, designers are turning to near-data processing (NDP), which places compute units closer to memory to reduce data movement and latency~\cite{pimsurvay}. By processing data near or within memory, NDP can alleviate memory bottlenecks and improve performance, energy efficiency, and throughput.

Given these advantages, NDP has become particularly attractive for database applications (DB), which are inherently data-intensive and fundamental across various domains such as business intelligence, e-commerce, healthcare, and finance~\cite{garcia2009database}. 
Among DB operations, joins are especially critical, as they combine data from multiple tables or relations and require extensive memory access and processing of large datasets~\cite{plattner2012memory}. The performance of join operations is therefore closely tied to memory efficiency and plays a major role in determining overall query latency.
\begin{figure}
    \centering
    \includegraphics[
    width=0.7\linewidth]{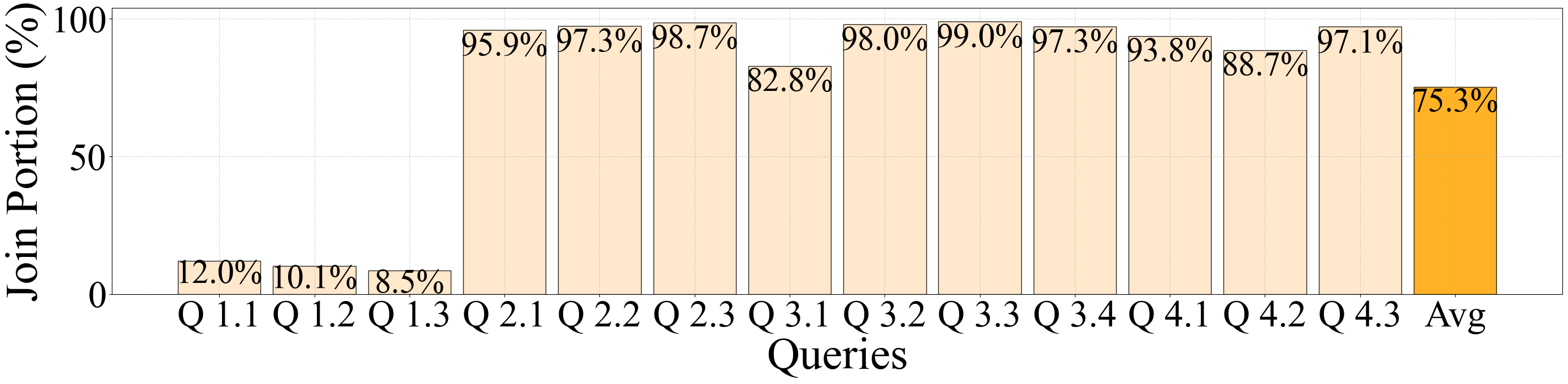}
    ~\caption{Percentage of join latency over total query computation for SSB benchmark~\cite{SSBQUERIES} using DuckDB~\cite{duckdbdocs} (Setup in Section~\ref{sec:Section 4.1}).
    }
    \label{fig:motivation}
\end{figure}
As shown in Figure~\ref{fig:motivation}, a substantial portion of query computation latency is attributed to performing joins. 

Joins in in-memory databases, despite eliminating disk I/O bottlenecks, face significant performance challenges.
Their performance is limited by memory access—CPUs stall on DRAM bandwidth saturation~\cite{cahceaccesskimsorthash,radixhashjoinHWoptimized},
exacerbated by growing core counts and SIMD widths, and the memory wall. While caches exist to bridge this gap, they often fail to benefit join operations due to the irregular, random access patterns leading to frequent cache misses and Translation Lookaside Buffer (TLB) thrashing, negating their intended speedup~\cite{radixhashjoinHWoptimized,joincacheproblem}. Non-Uniform Memory Access (NUMA) effects further complicate matters, as accessing remote memory can reduce throughput by up to 85\% compared to local access, necessitating NUMA-aware designs~\cite{pmpsortjoin,lang2013massively,numafriendly,cahceaccesskimsorthash}. 
 Parallelism, used to speed up computation time, 
 introduces additional complexity, requiring thread coordination and incurring synchronization overhead when building and probing shared data structures~\cite{lang2013massively,balkesen2013multi}. 
 Data skew compounds these problems, creating load imbalances where some threads process more keys while others are idle, hampering scalability and requiring techniques that add extra complexity~\cite{balkesen2013multi,chen2024skew,hardwareinmemoryalgoskew}. 
 While algorithmic solutions don’t fully resolve these issues~\cite{chen2024skew,hashjoin2,PIMJOINDRAM,hardwareinmemoryalgoskew}, they add overheads such as random writes and $\Omega(n \log n)$ complexity, making performance hardware- and dataset-dependent~\cite{balkesen2013multi,cahceaccesskimsorthash}. Figure~\ref{roofline} illustrates these challenges using a Roofline analysis of join execution in DuckDB~\cite{duckdbdocs}, one of the fastest existing database management systems (DBMS) for CPU-based join processing. This tight interplay between algorithm and architecture underscores the need for solutions that balance memory, algorithms, and hardware optimization.

To mitigate the memory wall effect on join and improve DB operation performance, prior work has explored hardware-based and NDP join acceleration~\cite{PIMJOINDRAM,SPID,bloompim,HMCjoin,pmpsortjoin,mirzadeh2015sort,radixhashjoinHWoptimized}, particularly DRAM-based Processing-in-Memory (DRAM PIM), which integrates computation into memory subsystems.
Current research on DRAM PIM, employs CPU-based algorithms to enhance data access bandwidth~\cite{mirzadeh2015sort,PIMJOINDRAM,SPID}. 
However, this approach does not fully leverage PIM's inherent capabilities for parallel execution and data access, and suffers from degraded performance due to PIM limitations. 
CPUs are centralized processors with rich compute resources and memory hierarchies that hide high latency and limited bandwidth, and PIM has limitations in computational power and data transfer capabilities between memory cells, operating differently from CPUs~\cite{pimsurvay}. As a result, algorithms that are optimized for CPUs may not work or suffer from limited performance when adapted to PIM~\cite{JOINPIMHCMALL} as they may fail to exploit the full potential of PIM's parallel data access and execution capabilities, or require additional cross-chip data transfers.

{%
This paper presents JSPIM, a novel PIM module designed to revolutionize database queries, with a particular focus on JOIN and SELECT operations. Performance of joins is critical to nearly every analytical query~\cite{hashjoin2} (see Figure~\ref{fig:motivation}), while SELECT is the most frequently executed operation in both database workloads~\cite{selectfrequent}. 
To summarize, database joins face four fundamental challenges:
\begin{itemize}
    \item \textbf{Memory-Bound Computation}: Memory access bottlenecks cause over 80\% of join query latency~\cite{fpgajoin}, with cache and TLB misses stalling pipelines~\cite{joincacheproblem}.
    \item \textbf{Algorithmic Complexity}: Parallelizing joins to mitigate memory bottlenecks adds overhead from synchronization, partitioning, NUMA effects, and I/O contention, often yielding minimal or negative performance gains~\cite{radixhashjoinHWoptimized,lang2013massively,balkesen2013multi}. 
    \item \textbf{Accelerator Limitations}: PIM cores suffer from limited compute capacity and costly cross-chip data transfers, hindering CPU-optimized join algorithms from leveraging PIM parallelism~\cite{PIMJOINDRAM}. GPUs suffer from host-device transfers and skew~\cite{chen2024skew,lin2019gpu}(Sec.~\ref{4.2.3}).
    \item \textbf{Data Skew}: Skewed data distributions and duplicate values degrade join performance, with hash joins being particularly susceptible~\cite{chen2024skew,duggan2015skew}.
\end{itemize}}
{
\begin{figure}
    \centering
    \includegraphics[trim={2.5cm 2.9cm 2.3cm 1.3cm},clip,width=0.9\linewidth]{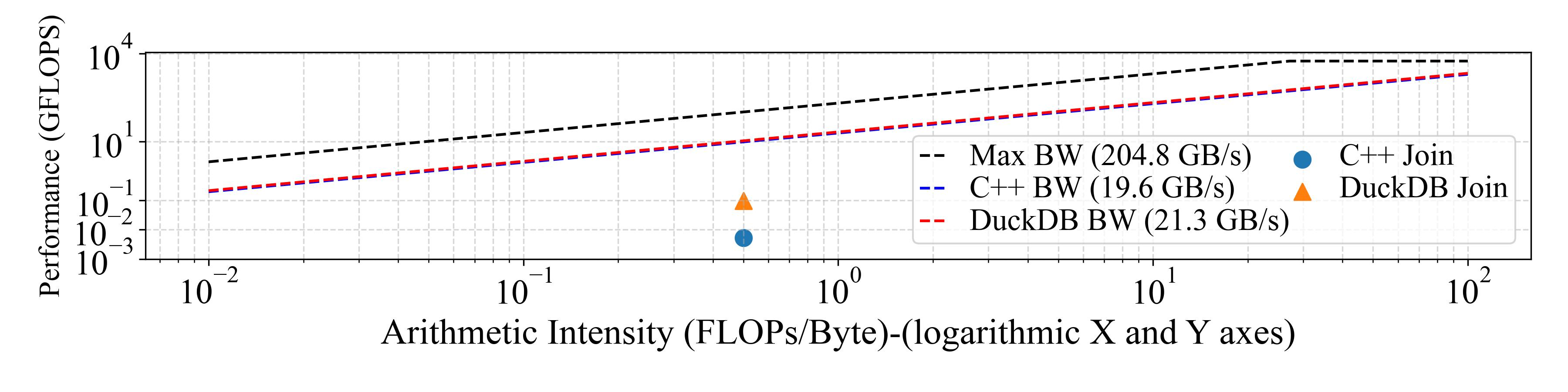}
    \caption{Roofline of SSB SF100 DuckDB join on CPU (setup in Sec.~\ref{sec:Section 4.1}).}
    \label {roofline}
\end{figure}
JSPIM accelerates hash join operations on Load-Reduced DIMM (LRDIMM)-based PIM to overcome memory access bottlenecks in database workloads. LRDIMM provides a scalable, cost-effective alternative to other PIM architectures by avoiding their thermal and integration challenges~\cite{chameleon}. JSPIM reduces data movement overhead by combining hardware and algorithmic innovations that exploit parallelism within the memory subsystem.}
{
To reduce algorithmic complexity and cross-chip communication, JSPIM builds the hash table once and stores it directly in PIM memory, eliminating repeated construction and minimizing data movement. Each PIM rank is equipped with a Rank-Level Unit (RLU) that operates on a compressed copy of the join-key column distributed across ranks. The RLU traverses table entries locally, bypassing the CPU involvement on match finds and exploiting rank-level parallelism. Implemented as a pipelined unit, the RLU enables parallel traversal, comparison, and communication, which improves throughput and avoids traditional synchronization overhead.}
{
JSPIM introduces subarray-level comparators that perform in-place hash table lookups, eliminating expensive CPU–memory data transfers. Unlike CPU-based designs—limited by sequential bucket traversal, cache capacity, and memory latency—JSPIM supports large hash buckets with fine-grained subarray-parallel scanning. Each bucket is mapped to a subarray row, and integrated hardware comparators examine all entries of a selected bucket simultaneously, enabling constant-time match detection even for large buckets (e.g., 200 entries). In contrast, ideal CPU-based hash tables assume bucket sizes of one to achieve true O(1) lookup,\label{onevslarge} which is rarely feasible in practice~\cite{hashjoin1}. To handle duplicates without inflating the hash table size, we build a duplication linked list to keep duplicate keys out of PIM and store unique key-value pairs in the hash table. This ensures predictable latency and avoids CPU-style probing across largely skewed buckets. The design employs a simple, hardware-friendly hash function to distribute keys across large buckets, effectively mitigating skew without requiring costly rehashing. Further, the design uses lightweight encoding to reduce the PIM-resident dataset and aligns data layout with DRAM’s bank and column structure for fast access. Together, these techniques deliver scalable, low-latency join processing while avoiding the complexity and bottlenecks that limit CPUs, GPUs, and prior PIM-based accelerators.
In summary, this paper makes the following contributions:}
{
\begin{itemize}
\item \textbf{Subarray-level PIM parallel search engine}: Integrated hardware comparators in subarrays per rank scan large hash buckets in parallel, delivering constant-time 
search across all entries. This fine-grained parallelism, implemented in select DRAM chips per rank called PIM chip, ensures low hardware overhead and scalable, fast data access unaffected by dataset size.
\item \textbf{PIM Rank-Level processing with RLU}: The RLU of each PIM rank reads keys and controls the PIM chip, enabling pipelined, parallel traversal, and comparison locally. This eliminates CPU involvement and cross-chip communication, minimizing data movement and synchronization overhead while achieving high throughput.
\item \textbf{Synergized algorithm–hardware data design}:  A duplication linked list manages duplicates, storing unique data in the hash table to prevent duplication-induced skew and ensure predictable latency. A simple hash function distributes keys across large buckets, mitigating skew effects without expensive rehashing.
Lightweight encoding minimizes the PIM-resident dataset, while data alignment with DRAM’s bank and column structure enables fast access.
\end{itemize}
}
We compare JSPIM with a CPU-based system using DuckDB~\cite{duckdbdocs}, a modern in-memory analytical database, on a PowerEdge R750 equipped with an Intel Xeon Gold 6330 CPU and 32 DIMMs of DDR4 3200 MHz DRAM memory in addition to 200GB intel optane NVMe as swap space. 
JSPIM is simulated using DRAMsim3 with DRAM settings identical to the CPU system (DIMM DDR4 3200 MHz). Performance evaluation is conducted using the Star Schema Benchmark (SSB), derived from the TPCH workload~\cite{SSB}. We first measure SSB join and scan queries latency on databases of varying sizes to evaluate the performance improvements for individual operations. Subsequently, we integrate JSPIM into DuckDB to assess the end-to-end query computation performance gains.
To compare JSPIM with state-of-the-art PIM solutions, we evaluate join performance using generic tables with 32-bit keys and values on a system featuring four memory channels and 16 PIM ranks. JSPIM’s performance is benchmarked against PID-Join~\cite{PIMJOINDRAM} and SPID-Join~\cite{SPID}.

{Our results demonstrate that for join queries, JSPIM achieves a range of 400x to 1000x speedup over DuckDB~\cite{duckdbdocs} running on CPU and a range of 76x to 870x speedup over NVIDIA RAPIDS~\cite{nvidia_rapids} running on A100 GPU and a range of 15x to 300x speedup over SPID-join~\cite{SPID}, a state of the art PIM design. Combining DuckDB with JSPIM for full SSB benchmark~\cite{SSB} query execution results in speedups ranging from 1.1x to 28x for individual queries, with a geometric mean speedup of 5.7 (cold) and warm (2.9). Comparing the execution time of the entire query flight (sum of all the query times), JSPIM's speedup relative to DuckDB is 2.5x.  These gains come at a modest cost, with JSPIM incurring only 7\% data overhead and 2.1\% area overhead per DRAM chip per rank.}
\section{Background}
This section summarizes relevant details on DRAM memories and PIM architectures, and introduces key join query algorithms in databases.
We also discuss state-of-the-art PIM designs supporting joins, identifying their limitations and enhancement areas.
\subsection{Dynamic Random Access Memory and PIM}
Dynamic Random Access Memory (DRAM) is currently the dominant memory technology in computing systems. 
Successive generations of DRAM technology (DDR2 - DDR5~\cite{mem}) offer improved transfer rates, 
power efficiency, and capacity.
DRAM modules~\cite{mem}, such as Dual In-Line Memory Modules (DIMMs), interface between the memory controller and DRAM chips. DIMMs come in various form factors and chip configurations. 
Advanced forms include LRDIMMs (Load-Reduced DIMMS), which incorporate an additional per-chip buffer between the pins and the data array, reducing electrical load on the memory controller and enabling larger memory configurations with improved signal integrity. These modules, supported by DDR4 technology, are ideal for memory-intensive applications and server environments requiring high capacity and performance~\cite{DDR4}.
  
Within DIMMs, memory chips are organized into ranks - logical groups of DRAM chips accessed as a single entity. 
While multiple ranks contribute to overall memory capacity and performance, in normal DRAMs, CPUs can access only one rank per DIMM in parallel~\cite{rankparalell}. 
Banks are subdivisions within a rank that contain groups of memory cells~\cite{memorybook}.
Each bank can be accessed independently, allowing for concurrent access to different areas of memory~\cite{microndatasheet}. 
Each bank in a DRAM chip contains multiple subarrays~\cite{Lisa}. A subarray 
refers to a group of memory cells sharing 
bit lines, organized into rows and columns~\cite{memorybook}. 
Each subarray has a row buffer 
used as temporary storage during read and write operations to get faster access to subsequent data within the same row~\cite{memorybook}.

Processing-in-Memory (PIM) is an innovative computing architecture that integrates processing elements directly into memory subsystems, enabling computation and data processing tasks to be performed within memory modules. This approach minimizes data movement between CPU and memory, enhancing overall system performance and efficiency. PIM architectures leverage memory systems' parallelism and high bandwidth capabilities to accelerate various applications, including database queries, machine learning algorithms, and scientific simulations. PIM implementations vary based on the memory technology used, with High Bandwidth Memory (HBM) PIM integrating processing elements into HBM stacks, but facing power management and scalability challenges in 3D stacked technology. These limitations particularly impact memory-intensive workloads that demand large memory capacity~\cite{chameleon}. Alternatively, DDR DRAM PIM architectures incorporate processing elements into DDR DRAM modules, typically presenting fewer thermal challenges than HBM-based PIMs, making them easier to construct. Some DDR-based PIM devices, such as UPMEM DIMMs, are already commercially available as PIM-enabled DIMMs~\cite{upmempim}.

When designing a PIM device, strategic placement of processing elements within the memory hierarchy is critical, balancing speed with overhead. As we move from memory cells to DIMMs, bandwidth, performance, and parallelism decrease, along with overhead and power usage. UPMEM PIM \cite{upmem2019} adds processing units to each bank for general-purpose applications. Key additional challenges in PIM design include integration complexity and scalability, ensuring compatibility between components, and supporting various problem sizes. 
Specifically, implementing hash join algorithms in PIM faces challenges due to memory limitations when accessing entire tables, necessitating CPU involvement and causing performance degradation. Data movement within PIM introduces latency, potentially offsetting integration benefits. The slower clock rate of PIM compared to CPU eliminates advantages in case of the need for sequential processing. These factors contribute to decreased performance of join computations using PIM compared to CPU\cite{JOINPIMHCMALL}. 
Addressing these issues is critical for using PIM to effectively enhance DB operations performance.

Nevertheless, recent work shows that well-designed PIM architectures can yield massive speedups on suitable workloads. For instance, \cite{pimlargeimp} reports up to 608x acceleration of full TPC-H queries using PIM for non-join operations, highlighting the benefits of minimizing data movement and tightly coupling compute with memory.

\subsection{Join Algorithms}
\label{sec-join-algorithms}
Relational joins, the key operation for data analytics, combine rows from two or more tables based on a predicate. The main types of joins include inner joins, outer joins (left, right, full), and cross joins. 
Database management systems rely heavily on join queries to combine data from multiple tables based on shared attributes, providing meaningful insights\cite{dbbook}. Among the algorithms for performing joins, hash joins are particularly popular due to their efficiency. In a hash join, a hash table is created for the smaller table, and used to find matching entries via a scan of the larger table. Various versions of the hash join algorithm have been developed, including partitioned hash join for parallel processing, Grace hash join for handling large datasets with limited memory, and hybrid hash join, which combines features of hash and nested loop joins~\cite{dbbook}.

Unfortunately, data skew causes uneven  distribution among hash buckets and undermines hash join performance ~\cite{perfecthash}. 
Robust hash functions, however, require prior dataset knowledge and are not always feasible. To mitigate the effects of data skew on performance, adaptive and hybrid hash joins are proposed which use data partitioning,  multiple hash functions, or  hash table resizing~\cite{perfecthash}.

The primary bottleneck in traditional hash joins is the extensive data movement between the CPU and memory, which can hinder performance and scalability, especially with large datasets. 
GPUs often outperform CPUs due to their high degree of parallelism, effectively exploited by many hash join algorithms \cite{lin2019gpu}. Studies such as~\cite{lin2019gpu} demonstrate that GPUs can process hash joins faster than CPUs under favorable conditions.
However, GPU performance advantages depend on factors like query complexity, particularly the number of join conditions and subsequent operations; data locality and transfer patterns, including whether data needs to be repeatedly moved between host and device memory; and data skew levels, which can impact load balancing across GPU threads more severely than CPU cores. 
If these factors do not align well, CPUs may perform similarly or better.
Recent work by Shimin Chen et al. \cite{chen2024skew} delves into this balance, proposing skew-conscious hash join algorithms  optimized for highly skewed data. Their research improves state-of-the-art hash join execution models on both CPUs and GPUs, showing that both systems perform hash joins well under conditions and CPUs perform efficiently in many scenarios.

PIM architectures offer a promising solution by integrating processing units directly into memory modules, reducing the need for frequent data movement between CPU and memory. Implementing hash joins in a PIM environment could enhance efficiency by minimizing data transfer latency, aligning with the trend of reducing data movement to improve processing speed and efficiency in computer architecture.
\subsection{Join Query Processing Using PIM}
Studies suggest that PIM can enhance join computation performance, contrary to Kepe's findings \cite{JOINPIMHCMALL}. 
Lim et al.~\cite{PIMJOINDRAM} tackle the memory wall challenge using UPMEM, a PIM architecture with a rank-level design. Among hash join, sort-merge, and nested-loop join algorithms, hash join proved the most suitable for the PIM architecture used in their study. The paper refines the partitioned hash join algorithm to account for the limited capabilities of scalar In-DIMM Processors (IDPs). It partitions fact and dimension tables across chips, enabling each processing unit to construct hash tables based on the dimension table chunks. To enhance parallelism, the fact table data is reorganized, grouping related data within the same chip. Additionally, a bank-level DRAM Processing Unit (DPU) uses registers for temporary data storage, eliminating CPU involvement during data transfers between chips. The matching results are then sent to the CPU for storage in main memory~\cite{PIMJOINDRAM}.

While this architectural approach improves join query efficiency compared to CPU-based processing in some cases, it has notable limitations. Inter-DRAM chip communication introduces a performance bottleneck, reducing PIM parallel processing advantages. Dedicated processing elements for each DRAM chip are primarily engaged in communication tasks, resulting in increased latency and poor parallelism. Hash tables larger than limited scratchpad capacity are not supported. The design is also sensitive to data frequency in the fact table, causing imbalanced distribution among PIM units and increased latency due to synchronized operation constrained by the slowest unit, with the impact from skew being greater than that when running on a CPU. Large fact tables exceeding PIM size require multiple join queries from the CPU, decreasing computation performance. These factors contribute to scalability issues and poor resource utilization in overall query computation.
To mitigate data skew caused by varying data frequencies in the fact table, SPID-Join~\cite{SPID} replicates join keys across banks and ranks, but this incurs high data overhead. Inter-rank transfers require CPU intervention, further increasing overhead, especially as word size grows.  
Additionally, SPID-Join requires careful tuning of the replication ratio, increasing complexity and causing performance degradation due to a larger memory footprint and higher inter-rank traffic. Unlike approaches that leverage duplication to reduce memory activations in PIM, SPID-Join does not exploit this advantage. Instead, it is constrained by limited WRAM capacity and relies on sequential hash searches. Furthermore, SPID-Join sends results to the CPU at the end, and excessive CPU-mediated inter-rank replication introduces bottlenecks, further limiting scalability.

Mirzadeh et al.~\cite{mirzadeh2015sort} compared radix hash join \cite{radixhashjoinHWoptimized}  and P-MPSM~\cite{pmpsortjoin} on CPU and HMC-based~\cite{hmc} PIM, finding sort-merge join more efficient. However, the algorithms weren't optimized for PIM's advantages.
The implementation of Bloom join on HMC-based PIM~\cite{bloompim} enhanced conditional join performance but did not extend to handling data skew and complex join queries like star joins. Star joins, a key operation in data warehousing, connect a central fact table to dimension tables in a star schema, enabling efficient, read-intensive analytical processing~\cite{dbbook,SSB}. While the study preserved algorithmic optimizations suited to traditional CPU and memory hierarchy, further tailoring these approaches to exploit PIM's capabilities could reduce the load and latency in the Bloom filter construction phase.
Research on HMC PIM for scan and join queries~\cite{HMCjoin} used partitioned hash-join but made unrealistic assumptions about data matching, potentially leading to challenges with repetitive word existence, uneven data distribution, and the data skew problem in real-world scenarios.
A comparative study~\cite{JOINPIMHCMALL} found that while selection and projection operators in join suit PIM, aggregation performs better on x86. Importantly, implementing join solely on PIM increased execution time, highlighting the need for algorithmic or hardware adaptations to leverage PIM effectively. Another problem associated with using HMC-PIMs is related to scalability, thermal and power management, which is addressed in  Chameleon~\cite{chameleon}. Chameleon introduces the possibility of putting accelerators in the data buffer devices of LRDIMMs, eliminating the need for expensive 3D-stacking technology and improving the bandwidth and data access throughput compared to other methods that aggregate accelerators near the CPU~\cite{chameleon}.

In summary, while PIM shows potential for database joins, current approaches face challenges including inefficient data movement, poor scalability, execution imbalance due to the data skew,  and limited ability to handle large datasets. Further innovation in both PIM architectures and algorithmic mapping is needed to fully exploit the benefits of NDP for DB operations such as joins.

\section{JSPIM Design}
We present JSPIM, a PIM architecture to enhance the performance of \emph{join and select} queries. We begin with an overview of the architecture, which incorporates subarray-level and rank-level PIM to maximize parallelism. We then detail supported operations, highlighting the efficient mapping of join operations to PIM hardware.
\subsection{JSPIM Architecture}
\begin{figure}[h]
    \centering
    \includegraphics[
    width=0.6\linewidth]{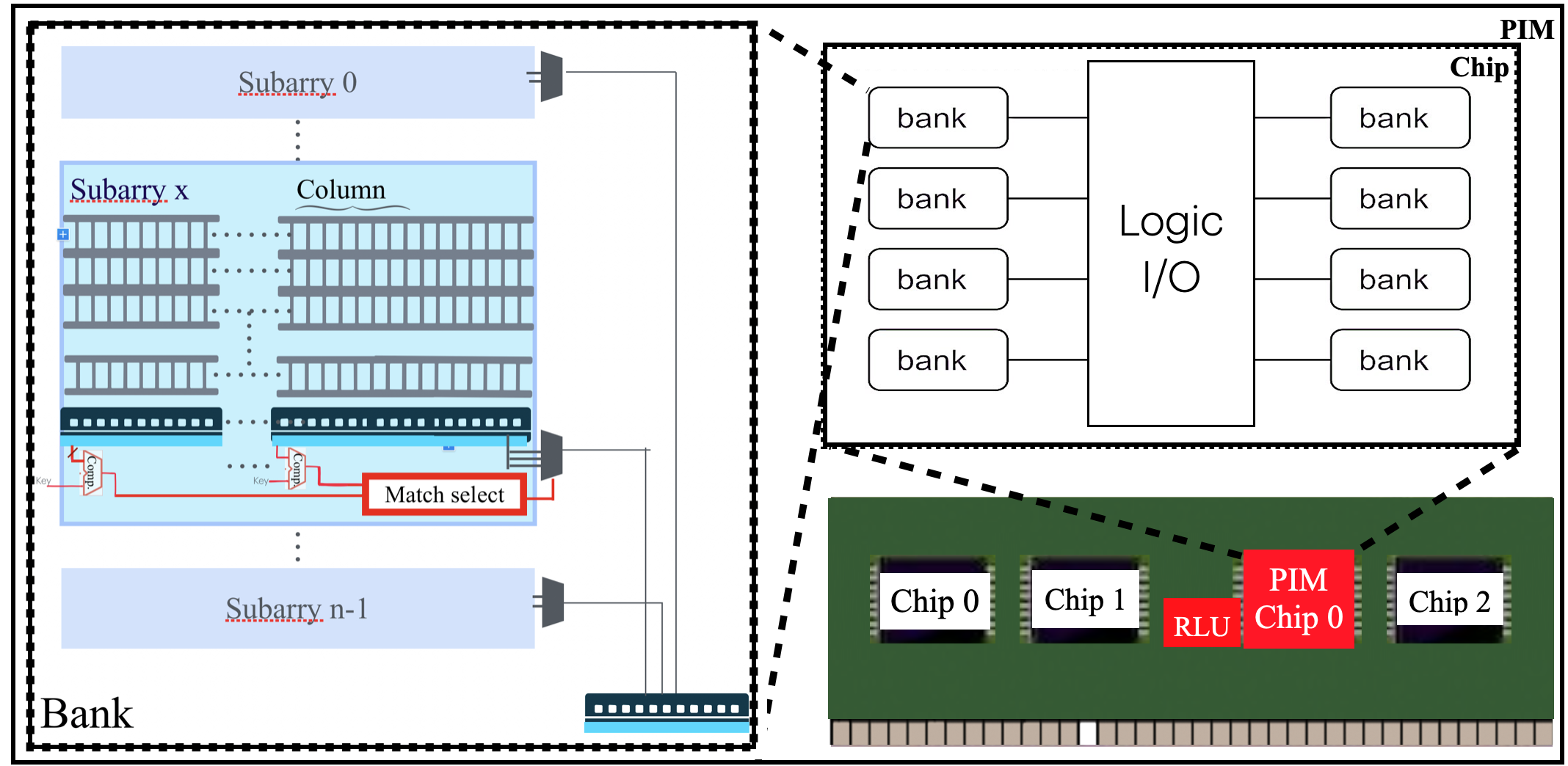}
    \caption{JSPIM Architecture. (Added modules shown in red)}
    \label{fig:overall}
\end{figure}
The JSPIM architecture is designed to operate flexibly as both traditional memory and a PIM system, with a mode-switching capability controlled through special address commands. The architecture introduces a Rank Level Unit (RLU) in each DRAM rank and adds subarray-level comparators to PIM-enabled chips. The architecture of JSPIM is shown in Figure~\ref{fig:overall}, with one chip armed with subarray-level hardware, called a PIM chip.
This section outlines the two primary components of the architecture: sub-array level elements and rank-level processor. 

\subsubsection{Subarray-Level PIM}\label{3.1.1}
JSPIM’s core search engine operates at the subarray level. Its main task is to take a key as input, search for it in the dataset, and return the associated value if found. If the key is not present, the module outputs a null value. To implement this, certain chips are designated as PIM-enabled and equipped with search engine elements. These PIM-enabled chips are used specifically to store the search dataset.
\begin{figure}
    \centering
    \includegraphics[
    width=0.6\linewidth]{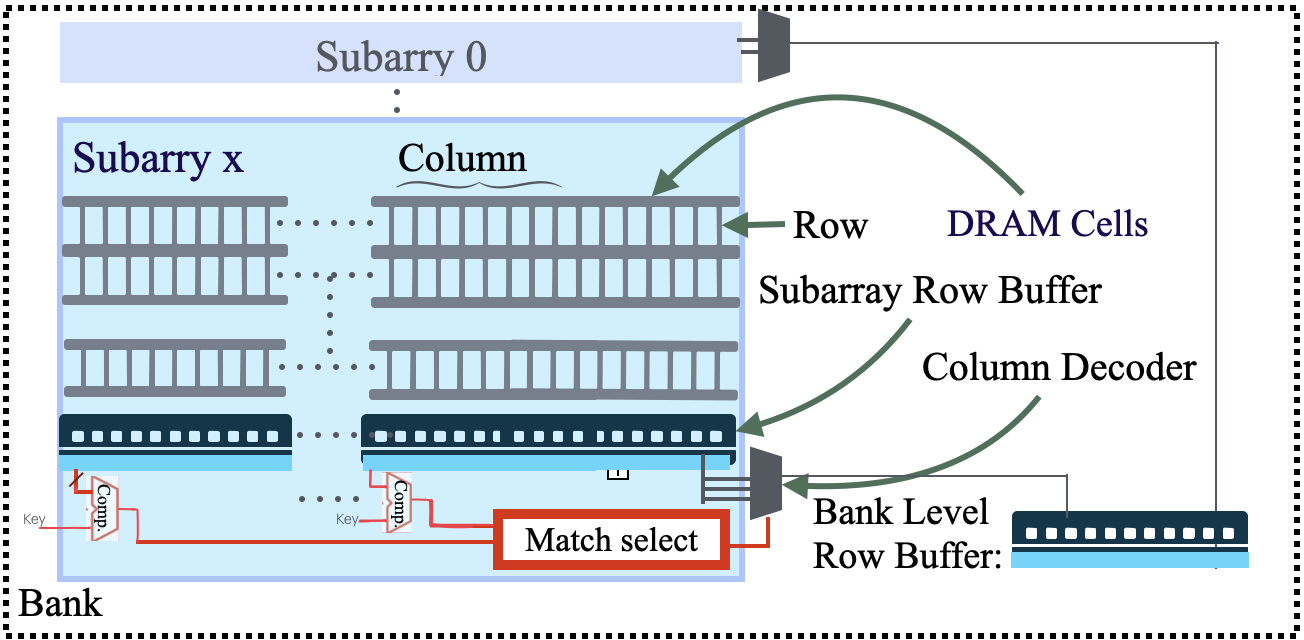}
    \caption{Subarray Level Architecture Inside a PIM Chip. (Added components 
    shown in red).}
    \label{fig:subarray}
\end{figure}

In each PIM-enabled chip, a set of comparators is placed after the row buffers within the subarray. These comparators perform parallel comparisons between a reference (probe) key and keys fetched from the row buffer. Each comparator is connected to bits corresponding to a single key in the row buffer and the reference key bits. Each key in the subarray is paired with a value that can be sent as output if a match is found.

When a comparator detects a match, it outputs a 1; otherwise, it outputs a 0. These outputs are then processed by a Match Select module, which encodes the comparator results to locate the address of the matched value. If no match is detected, the Match Select unit sets the result to null. 
JSPIM's subarray-level design is illustrated in Figure~\ref{fig:subarray}. The components highlighted in red represent the additional elements introduced by JSPIM to the baseline DRAM chip.

{
Implementing the search engine at the bank level reduces the number of comparators but introduces significant performance challenges. The smaller buffer size of banks in DRAM, typically 8 bytes, which is much smaller than the subarray-level row buffer (f.g, 1024 bytes) ~\cite{wang2020figaro} reduces parallelism when keys and values are stored together. Consequently, multiple key fetches would be required to read large buckets.
Stopping the searchIf the operation stops upon finding the relevant key, it results in varying latencies and sensitivity to data skew. On the other hand, reading the entire bucket for consistency incurs unnecessary overhead and performance degradation.
Additionally, this approach This approach complicates the match-select unit and introduces serial operations, increasing latency.}

\subsubsection{Rank Level Unit (RLU)} 
\label{RLU}
JSPIM uses an RLU to eliminate cross-chip communication by avoiding PIM to CPU communication. The RLU is a small processing unit that performs basic arithmetic operations. 
The CPU controls the RLU by sending write requests to specific predetermined addresses in DRAM. These commands include operations like PIM start, PIM off, and other supported queries. When a PIM start command is issued, the RLU operates in active mode, controlling the PIM chip. Conversely, sending the PIM off command switches the module to function like standard DRAM.
The RLU has access to LRDIMM's per-chip input and output data buffers, and can read the commands sent by the host CPU from these buffers as outlined in the Chameleon~\cite{chameleon} system. 
The PIM chip can operate both in regular memory mode and in PIM mode. 
{
When the CPU sends a write operation to the designated special address, the RLU recognizes the command by reading the input and switches to PIM mode, taking over and signaling the PIM chip.
}

{PIM designs such as UPMEM~\cite{upmem2019} place processing units within each DRAM chip.
In contrast, JSPIM designates specific DRAM chips as PIM-enabled while keeping others as regular DRAM chips. 
This design choice allows parallel/pipelined access to both PIM-enabled and regular chips within the same rank with the help of the RLU. This configuration enhances the efficiency of concurrent operations on the fact table and hash dataset during join operations by reducing communication over the memory channel.} 

{
Once in PIM mode, the RLU waits for the CPU/memory controller or Direct Memory Access (DMA) engine to send read requests for addresses, from which keys will be retrieved. The address is used to index/access the regular (non-PIM) DRAM chips, in which the keys are stored. These keys are cached in buffers within the RLU and used to drive the search on the PIM-enabled chip. 
The RLU drives the relevant row in the PIM-enabled chip to perform an associative search for the key by providing the reference key to the comparators.
The resulting values output by the search on the PIM chip can be 8 to 64 bits, depending on the Burst Length (BL) configuration specified in the Micron datasheet~\cite{micron_ddr4} (e.g., 8-bit corresponds to BL=1).
} 

{
The chameleon~\cite{chameleon} paper introduces three different methods of RLU integration. In chameleon-d the RLU uses the Register Clock Driver (RCD)~\cite{JEDEC_DDR4RCD02} BCOM pins~\cite{chameleon, JEDEC_DDR4RCD02} to send the address to DRAM chips~\cite{chameleon}. These pins are shared among all DRAM chips so we cannot send different commands to multiple chips using BCOM pins. In this architecture, our single RLU can be integrated into the RCD, but the access to regular DRAM chips will be serialized with access to the PIM-enabled chip, reducing parallelism and preventing pipelining. 
In chameleon-t and chameleon-s, the RLU uses per-chip DQ pins~\cite{chameleon, JEDEC_DDR4RCD02} to address the chip. These DQ pins are connected to dedicated data buffers for each chip of LRDIMM DRAM.  By using DQ pins for addressing, DRAM chips can be individually addressed.}

{The functionality of the RLU (iteration over keys) could potentially be integrated into a memory controller or DMA engine, which must then serially address DRAM and PIM over the memory channel, thereby reducing performance. 
Additionally, the CPU must assist in feeding keys from other chips to the PIM chip, leading to further inefficiencies.}

\subsection{JSPIM Operations} 
To perform operations, read and write requests are sent to PIM-DRAM. In this study, we assume the device is in PIM mode from boot time, bypassing the software overhead of virtual address management to concentrate on query processing latencies. 
JSPIM is primarily designed for join queries, and it also supports \texttt{select where(=)} and \texttt{select distinct} queries. The architecture includes update commands to keep the dataset synchronized over time. 
\subsubsection{JOIN Query}\label{3.2.1}

Hash join algorithms are fast, widely used, and PIM friendly. JSPIM selects the classic hash-join algorithm for several key reasons.  Since PIM is not limited by CPU cache size and therefore does not need to generate small hash tables to reduce data access latency, we do not use the partitioned hash join. Furthermore, the partitioning phase in other hash join algorithms (e.g., partitioned hash joins) introduces cross-chip communications and data replacements,
making them unsuitable for PIM architectures.

Unlike CPU-based algorithms that need to minimize bucket size for performance optimization, JSPIM's architecture \emph{supports large hash buckets} and prioritizes a \emph{straightforward hash function} instead of complex designs focused on collision reduction. This approach uses fewer PIM buckets, reducing row activation and power usage while increasing performance.

JSPIM addresses the Data Skew problem, caused by hash collisions and data duplication, through two strategies: 
constant-time in-bucket searches for large buckets and 
data compression to standardize data sizes across buckets in the absence of duplications. To handle duplicates efficiently, JSPIM maintains a duplication list separate from the main hash table, allowing parallel processing of duplicated data while the PIM handles the hash table operations. 

\begin{figure}
    \centering
    \includegraphics[
    width=0.7\linewidth]{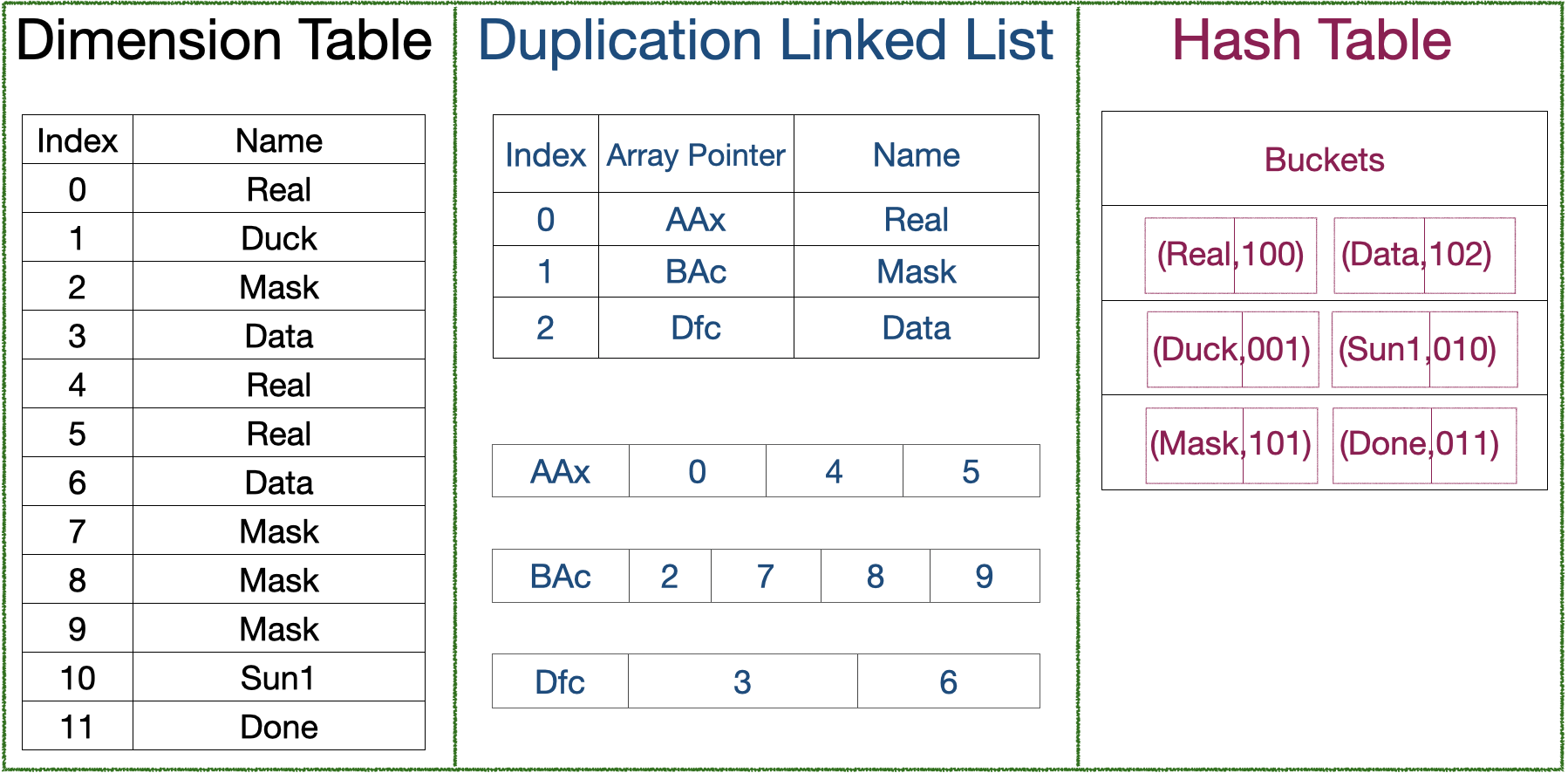}
    \caption{Sample Duplication Linked List and Hash dataset.} 
    \label{duplication}
\end{figure}

JSPIM ensures constant-time search results through a two-part strategy: constructing hash tables with unique entries and implementing a duplication-linked list (duplication table). This design separates duplicate data from the main hash table by storing a single key-value pair for each unique entry, with the value pointing to a duplication table entry that contains indices for all duplicate occurrences.  Algorithm~\ref{duplicationbuild} outlines the construction of the duplication table, and Figure~\ref{duplication} illustrates an example of a duplication table integrated with a hash table for a join query. In this figure, the hash table contains the unique keys from the dimension table and the relevant value bits hold the index information for the corresponding dimensional table row. Here, we added one bit to each value to indicate whether the data is from dimension table or the duplication list. This extra bit indicates whether the word in the hash table is unique or has replicas. The duplication-linked list holds pointers to multiple lists, each containing indices of the word's replicas.

For supporting joins across multiple columns, JSPIM offers flexibility by allowing either separate data sets, distinct hash tables in different ranks, or merged tables containing data for all columns.
\begin{algorithm}
\caption{Duplication Link List Construction}
\begin{algorithmic}
\REQUIRE Base table $B$
\ENSURE Hash table $H$, linked list $L$
\STATE Initialize $H$ and $L$ to empty
\FOR{each entry $e$ in $B$}
    \STATE $key \leftarrow$ hash of $e$
    \IF{$key$ in $H$}
        \IF{$key$ new to $L$}
            \STATE $value \leftarrow H[key]$
            \STATE Insert $value$ at head of $L[key]$
            \STATE Update $H[key]$ to head of $L[key]$
        \ENDIF
        \STATE Append $e$ to end of $L[key]$
    \ELSE
        \STATE Add $key$ to $H$ with $e$
    \ENDIF
\ENDFOR
\end{algorithmic}
\label{duplicationbuild}
\end{algorithm}

JSPIM employs dictionary encoding to store fact table keys and hash table pairs within the PIM. Dictionary encoding is a data compression technique commonly used in databases, where frequently occurring values are replaced with shorter codes from a predefined dictionary to save space. In JSPIM, data values are stored as fixed-size codes, ensuring consistent storage usage per entry. During the encoding phase, the system can handle collisions within hash buckets by modifying the codes, enabling efficient data distribution across hash buckets. The decoding process is simple, involving just a lookup, which benefits from our optimized search engine.

To perform \emph{join lookups}, it’s crucial to determine how to store the encoded fact table entries and hash table in both PIM-enabled chips and DRAM. Since searches need to be conducted within hash buckets, we store the hash table within a subarray-level architecture in PIM. To maximize the performance of our search engine, we map the hash buckets to the rows of PIM, allowing for a single-cycle search by activating a single PIM row. This brings the relevant bucket into the row buffer for a fast lookup.
To achieve this, we split the key bits into two groups. The first group consists of index bits, which are used to locate the unique row in PIM. The remaining bits are stored within the corresponding columns of the row, connected to comparators. Each key is accompanied by a set of relevant value bits that are not connected to comparators.

In database systems, there are two primary data storage structures: row-store and column-store. A row-store organizes records with fields for each attribute, making it efficient for accessing individual records~\cite{dbbook}. In contrast, a column-store stores each attribute in a separate array, providing efficient access to multiple attributes across many records\cite{dbbook}. Since join operations primarily work with table columns, JSPIM adopts a column-store approach, storing coded Fact table entries (keys) in DRAM rows.

{Figure \ref{data} shows the data layout in PIM. In this example, each bucket holds two entries, but in real DRAM, it is about 100 times larger. Characters are stored as 1-byte binaries, shown in character format for clarity. Key bits used for indexing are not stored in the PIM chip 
{(shown in gray). }
{When accessing data in PIM mode,} standard DRAM chips (without subarray-level compute logic) 
{are controlled via command/address (C/A) signals sent to the PIM-enabled rank}, while the PIM chip’s addressing is managed by the RLU 
{(see ~\cite{chameleon}). The interface to the PIM-enabled rank remains unchanged. In order to initialize the data for the PIM computation, the software will need to combine data from two datasets (Fact and Hash table) when writing each word.}
The hash table is build using the Dimension table. The duplication list is in CPU memory. DB is the DRAM chip data buffer in LRDIMM.
\begin{figure}
    \centering
    \includegraphics[
    width=0.8\linewidth]{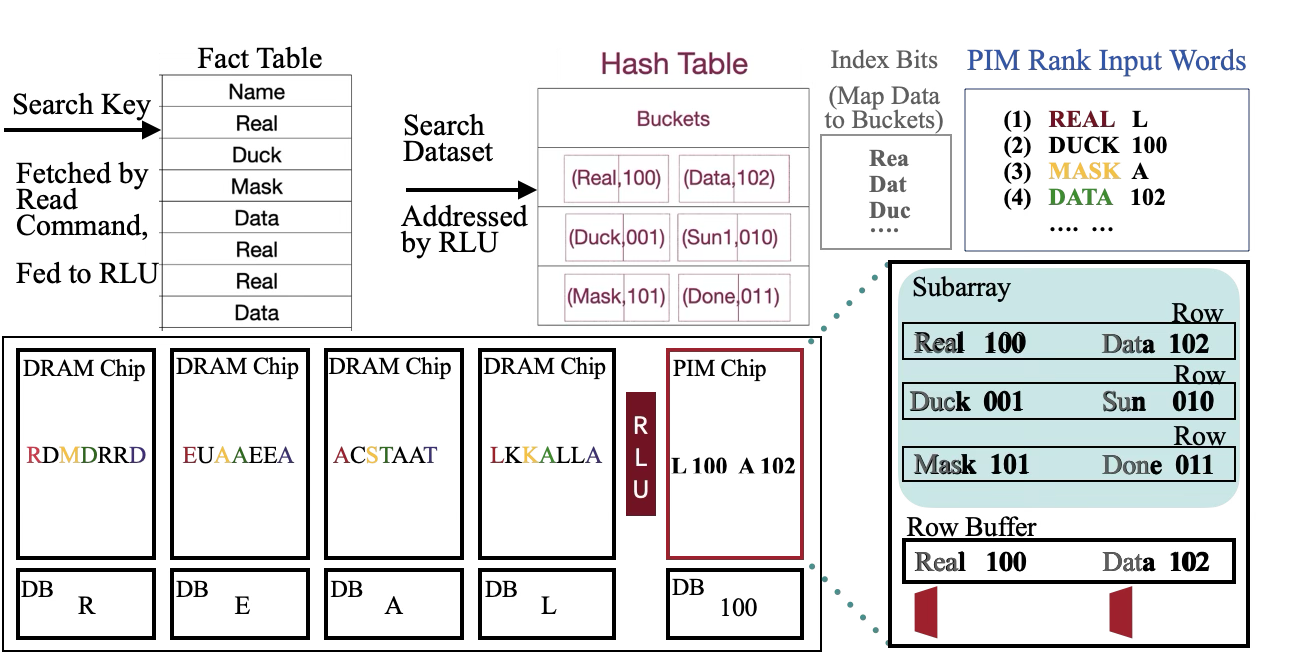}
    \caption{
    {Join Layout in PIM Rank’s DRAM/PIM Chips.} }
    \label{data}
\end{figure}
}
 \emph{CPU-PIM Interface.} To initiate a join operation, the application 
issues a stream read commands starting at the first fact table entry address of interest.
Each read request retrieves potentially multiple contiguous chunks of keys into the DRAM buffer, from which it is copied to RLU buffers.

\begin{figure}
        \centering
        \includegraphics[
        width=0.7\linewidth]{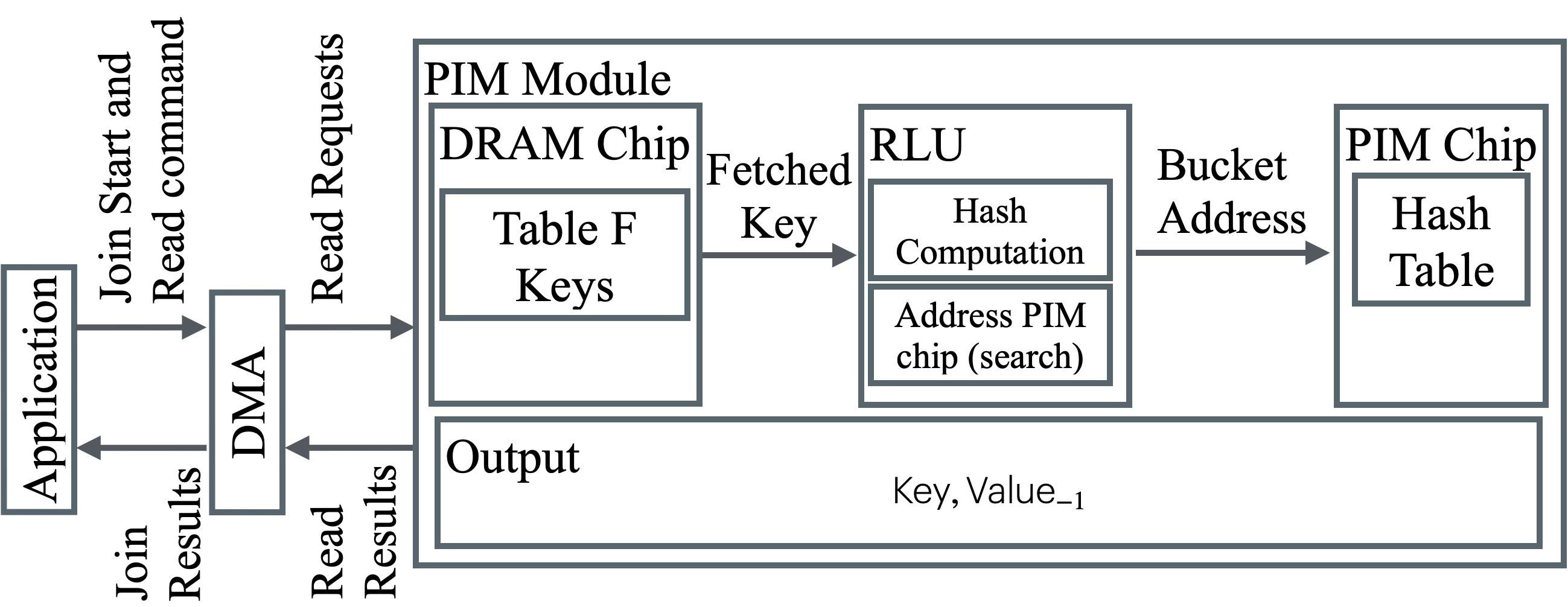}
        \includegraphics[
        width=0.7\linewidth]{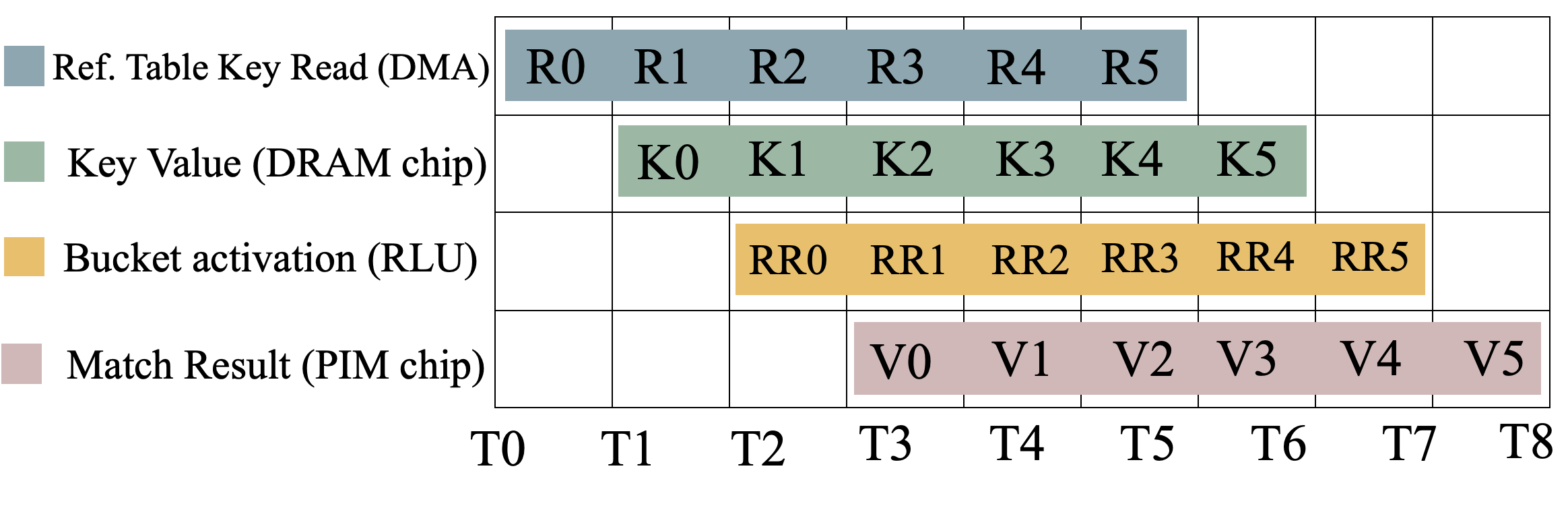}
            \caption{JSPIM Join Execution Pipeline. The diagram shows data flow, module commands, and a pipelined task timeline. While probe keys are fetched, result keys are sent in parallel.}
    \label{fig:pipeline}
   \end{figure}

The RLU uses pipelined parallelism by enabling key value reads from DRAM in parallel with key search within PIM, creating a pipelined workflow, where four operations occur concurrently: Reading key values from the reference table, copying the result to RLU Buffers; value search within the PIM chip; and return of the match result;  as shown in Figure~\ref{fig:pipeline}. 
To prevent buffer overflow in the RLU due to the mismatch of pipeline stages, the RLU must determine the necessary stall time: the number $N$ of responses from PIM before issuing further commands. The value of $N$ is influenced by factors including the RLU buffer size, burst length (BL), key bit width, number of DRAM chips per rank (DCR), and DRAM device size. 
Since the fact table is distributed across multiple DRAM ranks, all ranks can process data in parallel— further accelerating complex multi-table joins. Additionally, because the fact table entries are stored contiguously in PIM, address generation and data fetching can be efficiently offloaded to Direct Memory Access (DMA), keeping the CPU out of the loop when necessary. Also, when the CPU wants to fetch the join result for some determined fact table keys, it can feed them directly to the PIM. The CPU-PIM interface for these instructions leverages the AVX architecture for vectorized memory access~\cite{intelavx512}.

When the fact table exceeds PIM capacity, the processor has two options. First, it can load data into the write buffer, allowing the RLU to read from it and perform the join. Alternatively, the processor can pipeline the data into the PIM's DRAM chips, using a double buffering technique. 

To optimize performance, the CPU and RLU can collaborate to filter out redundant values and transmit only non-redundant values. This strategy prevents unnecessary data transfers when fact keys are repeated. The RLU is equipped with an optimization buffer that functions like memory controller coalescing, filtering out redundant requests within a window before sending. The RLU does not optimize redundant requests that fall outside this window and therefore does not filter them. In parallel, the CPU ensures that redundant data is not requested from PIM within the same window. The optimal window size is determined by the number of fact keys that can be prefetched ahead of their associated value results. This is feasible because multiple chips fetch keys in parallel, while a single chip fetches values—allowing several sequential keys to be read per value fetch, depending on word size.

\subsubsection{Select Query}\label{selectq}
The \texttt{SELECT} is a useful tool for retrieving data, allowing users to specify the exact information they need from a database. The \texttt{SELECT DISTINCT} clause helps remove duplicate entries, returning only unique values from a specified column.
For the reference column for which a hash table has been created, JSPIM can easily handle this query by returning all the key values in the hash table, since they are unique. 

Additionally, the \texttt{select} statement can include a \texttt{where} clause to filter results based on specific conditions. When the condition is equality (\textit{"\texttt{select * from} table \texttt{where} column0 = value;"}), the query 
{finds any occurrence of a word in a table column, returning only the matching values. For this query in JSPIM, the application simply sends the command, which costs a single DRAM read. 

\subsubsection{
{Update Commnads}}\label{update1}
{
Our design incorporates update commands for persistent storage and incremental maintenance of hash tables across queries. As a result, JSPIM avoids repeated reconstruction costs and amortizes setup overhead over multiple queries. Its prebuilt hash table and duplication list are created once and reused in-place, enabling low-latency query execution even in data-intensive scenarios. The DBMs with JSPIM treats its duplication lists and PIM‐resident hash tables and similar to any other auxiliary structure in an RDBMS (e.g., secondary indexes or materialized views).
 This section describes the supported commands in detail.}

\emph{Entry Update}
JSPIM uses entry updates, to synchronize the PIM dataset with the main database. This command updates a bucket within the hash table or an entry within the fact table. For fact table updates, it provides the entry address and new data, similar to a DRAM write command. To update a hash table, the application must provide the RLU with the data using write commands directed to a DRAM-specific address. This ensures real-time updates and synchronization for fast join and scan queries.

\emph{Index Update}
This command submits the key and a new value, prompting the PIM to search for the key and, on a match, update the value.
It is recommended that applications aggregate frequent updates and execute them as a single bulk update to improve efficiency and reduce multiple index update queries. The index update command also eliminates the need to store the hash dataset outside PIM, thereby minimizing overhead.

\emph{Table Update}
The table update command leverages DRAM's burst write functionality to optimize performance within PIM architecture. It requires the initial address, command bits, and a dataset for writing into PIM cells. Applications can update fact table entries, hash table entries, or both concurrently. With three distinct supported updates, this command is the fastest supported by PIM. Buffering frequent updates and issuing them as table update queries effectively optimizes performance.

\section{Evaluation}\label{sec:Section 4.1}
We outline our system setup and evaluation methodology used to compare the performance improvements of JSPIM over state-of-the-art and best CPU-based, PIM-based, and GPU-based solutions for join and select queries. 
We simulate the performance of JSPIM and compare its performance to that of DuckDB, a state-of-the-art in-memory database management renowned for its fast join computation using partitioned hash join, as well as a C++ implementation of the classic hash join algorithm.    
We also compare JSPIM to SPID-Join~\cite{SPID} and PID-Join~\cite{PIMJOINDRAM}, two implementations on a commercially available PIM, UPMEM~\cite{upmem2019}, and to GPU-base systems using NVIDIA RAPIDS~\cite{rapids2020}. 
Finally, we discuss the overheads associated with our design, including data and area overheads.

\subsection{Experimental Evaluation}
\subsubsection{Methodology}\label{methodology}
{
We simulate JSPIM using DRAMSim3 simulator~\cite{dramsim}, configured to model PIM-specific behavior such as subarray-level activation and trace-driven memory access. The RLU has an 8-entry optimization buffer which behaves like a memory coalescing window, filtering out duplicate memory requests. CPU coordinates with PIM, avoiding redundant data access within the same window.  Table~\ref{configurationtest} presents the configuration of our evaluation system. The first set of rows provides  server specification details, while the second provides simulator configuration, chosen to align with server specification.

\begin{table}
\caption{System Configuration Details}
\label{configurationtest}
\centering
\begin{tabular}{|l|l|}
\hline
\textbf{Component} & \textbf{Specification} \\
\hline
Server Model & PowerEdge R750 \\
CPU & 112 Intel(R) Xeon(R) Gold 6330, 2.00GHz \\
GPU & NVIDIA A100 , 40GB HBM2\\
Memory & Registered DDR4 3200 MHz \\
DIMMs & 8 channel \& 16 DIMMs \\
Swap space & 200 GB Intel Optane NVMe \\
Architecture & x86\_64 ( 32-bit and 64-bit op-modes) \\
Operating System & Linux (5.15.0-91-generic) \\
\hline
Simulation Tool & DRAMsim3 \cite{dramsim} \\
Memory & 8 channels of DDR4 3200 MHz \\
Rows and Columns & 65536 rows, 1024 columns \\
\hline
\end{tabular}
\end{table}

Our benchmark workload is Star Schema Benchmark (SSB)~\cite{SSB}, which includes a fact table (\texttt{lineorder}) and four dimension tables (\texttt{DATE, SUPPLIER, CUSTOMER, PART}). The SSB, derived from TPC-H, emphasizes star structures representative of real-world data warehouses. We generate datasets using the \texttt{ssb-dbgen} tool~\cite{SSBDBGEN} at scale factors (SF) ranging from 1 to 100, scaling all tables linearly, with the fact table growing from $\sim$6M to $\sim$600M rows. This scaling enables us to systematically evaluate the system's data volume sensitivity and scalability at realistic scales. 

In addition to enabling evaluation of individual queries such as joins and scans, SSB supports a suite of 10 representative queries that reflect a variety of analytical patterns. We evaluate these 10 queries at SF100. They are grouped into four categories: Q1.x (predominantly filtering), Q2.x (two-table joins), Q3.x (three-table joins with moderate-to-complex filtering), and Q4.x (multi-way joins involving dimensional hierarchies). This categorization ensures coverage of a wide range of query types common in decision support systems.

We compare JSPIM against three hardware baselines: a CPU-only system, a GPU-accelerated system, and state of the art PIM. The CPU and GPU experiments are run on the same server used for JSPIM’s host platform. For CPU, we use DuckDB~\cite{duckdbdocs}, a modern vectorized analytical DBMS optimized for in-memory columnar processing, with all its default optimizations enabled.
For GPU joins we use the NVIDIA RAPIDS library~\cite{rapids2020}, which uses CUDA~\cite{cuda2023} to execute dataframe operations in parallel on the GPU, minimizing data transfer overhead and maximizing throughput for large joins. To handle data spilling and parallelization, we integrate Dask~\cite{dask}, which partitions datasets into manageable chunks and supports dynamic task scheduling~\cite{dask2016,rapids2020}. For all systems, input tables are preloaded into memory before timing queries, to fairly isolate query runs. Also, the SSB benchmark data is static, and there is no data update during tests.

For the PIM baseline, we compare against SPID-Join~\cite{SPID} and PID-Join~\cite{PIMJOINDRAM}, two state-of-the-art join accelerators running on UPMEM hardware. These systems are evaluated on a real UPMEM-enabled server with 4 memory channels and 16 PIM ranks, identical to the setup in SPID-Join~\cite{SPID}. The JSPIM configuration is reset to match the SPID in the memory channels, ranks and timing. We model both in-memory join execution and result transfer to the host, using identical datasets and workloads across all platforms to ensure fair comparison.
Since PID-Join did not support SSB data types, we used synthetic join workloads with 32-bit keys and values, over tables $R$ and $S$, starting with $|R| = |S|$ and progressively doubling the size of $S$ across four tests. Each configuration includes multiple data types and Zipf factors (0, 0.5, 1.5, 2) to evaluate the impact of skew. We test three sizes for $R$: 0.5M, 8M, and 32M tuples~\cite{SPID}.
}

\subsubsection{Comparison with CPU-based systems} 
This section presents key results for comparing JSPIM to the CPU baseline running multiple join query types.

\emph{Star Join results:} 
To assess the relative contributions of algorithm and hardware to performance, we compare a classic single-threaded hash join implementation in C++ (JSPIM’s base algorithm) with DuckDB, both running on the same hardware. We run $\texttt{fact} \Join \texttt{dimension}$, where the dimension tables have unique keys, and duplication list is empty.
Figure~\ref{fig:improvemnt} shows the join latency results and speedups of DuckDB over C++ (up to 52×) and JSPIM over DuckDB (up to 1001×).
\begin{figure}
    \centering
    \includegraphics[
    width=0.75\linewidth]{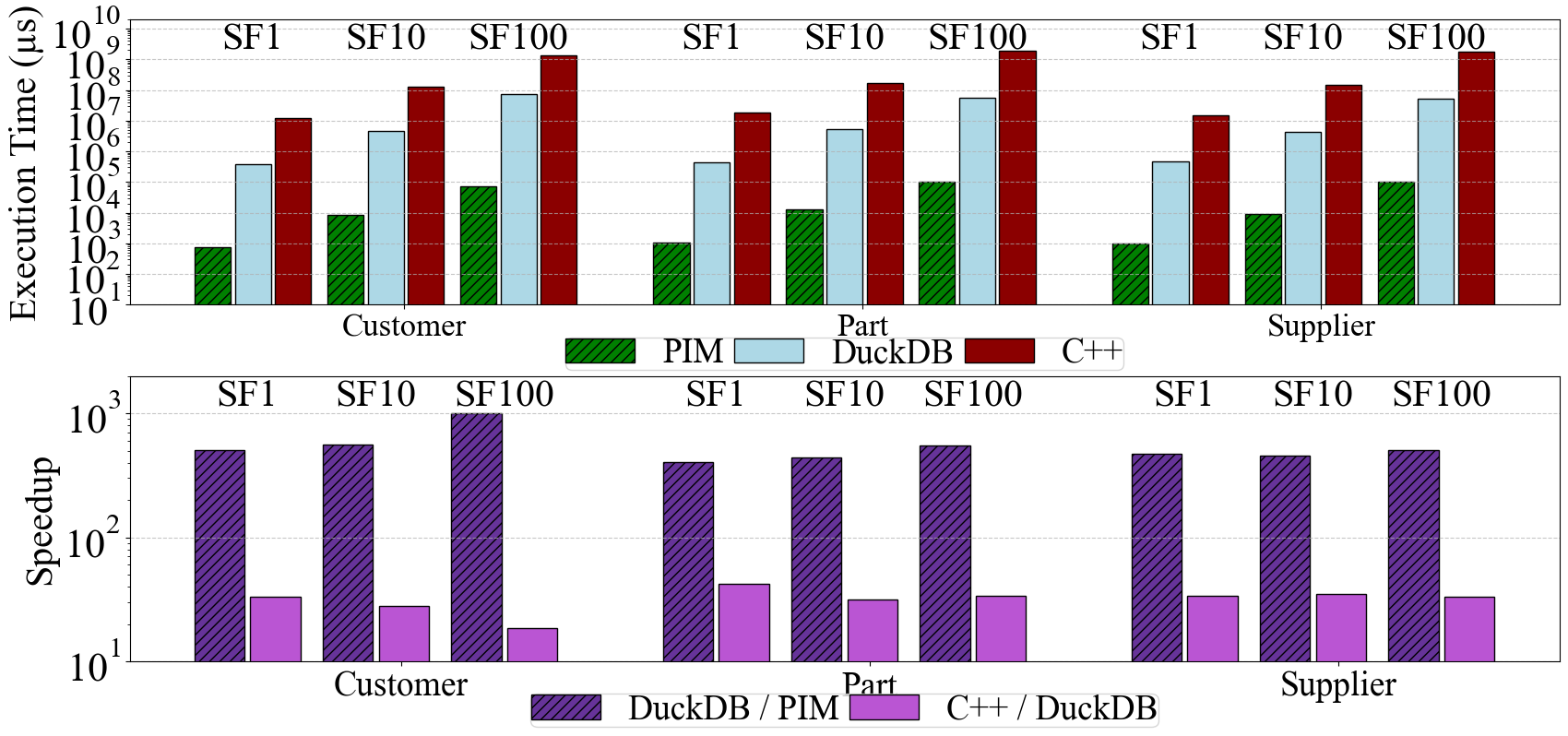}
    \caption{Comparison of JSPIM over CPU-based algorithms for join query {\footnotesize(e.g (\texttt{SELECT n.*, r.* FROM lineorder n JOIN part r USING (column);}))}. The duplication list is empty. (Y-axis in log scale).}
     \label{fig:joinlatency}
     \label{fig:improvemnt} 
\end{figure}

JSPIM’s speedup over DuckDB ranges from 400× to 1001×. For joins involving the customer and part tables, speedup improves with scale factor, while the supplier join shows a different trend due to key distribution variations affecting cache hit rates. The largest gains occur in the Supplier join, where the fact table contains approximately 200K repeated entries across 600M rows, with average and maximum consecutive repetitions of 3K and 3,220, respectively. This repetition reduces latency by minimizing PIM row activations.

{\emph{Setup phase cost results:}\label{setupphase}
Table~\ref{setup} presents the latency of hash table construction in both partitioned hash joins and JSPIM. To evaluate the setup phase of JSPIM, we measure the latency of two steps: (1) constructing and storing the hash table and duplication-linked list in memory (JSPIM Data Construction), and (2) populating the hash table and fact table entries into PIM (PIM Population). We compare this against the partition and build phases of the partitioned hash join.
For fair comparison, all phases are implemented in optimized single-threaded C++ using memory-mapped I/O.
To simulate PIM data population, we use the DRAMSim3 trace generator that sequentially writes data to PIM ranks~\cite{dramsim}. 

The latency of PIM population and PHJ partitioning for a given scale factor (SF) is dominated by the size of the fact table. However, for generating hash tables, the memory hierarchy and cache hits become more important. 
At scale factors 1 (SF1) and 10 (SF10), the \texttt{part} table incurs the highest data construction latency due to its larger number of entries compared to other dimension tables. Specifically, at SF1, the \texttt{part} table contains 200,000 rows (approximately 3.2\,MB with 16-byte entries), while at SF10, it scales to 2,000,000 rows (32\,MB). In contrast, the \texttt{customer} table has 30,000 rows (0.48\,MB) at SF1 and 300,000 rows (4.8\,MB) at SF10. However, at SF100, with the \texttt{part} table at 20,000,000 rows (320\,MB) and the \texttt{customer} table at 3,000,000 rows (48\,MB), a significant increase in hash table generation latency is observed for the \texttt{customer} table.

This latency spike for all dimension tables at SF100 arises when the hash dataset size exceeds the cache capacity (e.g., 32\,MB L3 cache), causing reduced cache hit rates and increased random memory accesses during hashing. The \texttt{customer} table’s key distribution exacerbates this issue by increasing collision rates in the hash table, amplifying collision resolution overheads. Unlike the \texttt{part} table, where random accesses dominate due to frequent interleaved accesses between fact and hash tables, the \texttt{customer} table’s latency is driven by both collision handling and random accesses. Although dimension tables like \texttt{part} and \texttt{customer} have no key duplication, the hash function’s mapping of actual data values introduces variability. 

\begin{table}
\centering
\caption{JSPIM Setup Latency Evaluation (ms).}
\label{setup}
\setlength{\tabcolsep}{2pt}
\renewcommand{\arraystretch}{0.95}
\begin{tabular}{p{2.7cm}p{2.6cm}p{2.2cm}p{4.4cm}p{2.7cm}}
\hline
\textbf{Table (SF)} & \textbf{PHJ Partition} & \textbf{PHJ Build} & \textbf{JSPIM Data Construction} & \textbf{PIM Population} \\
\hline
Customer (SF1)   & 979.25   & 13.5    & 5.2     & 0.1545 \\
Part (SF1)       & 1030.25  & 120.75  & 35.6    & 0.1545 \\
Supplier (SF1)   & 1007.0   & 0.8     & 0.3     & 0.1545 \\
\hline
Customer (SF10)  & 10056.5  & 151.5   & 56.4    & 1.6 \\
Part (SF10)      & 10210.5  & 581.0   & 152.0   & 1.6 \\
Supplier (SF10)  & 10092.75 & 8.0     & 3.0     & 1.6 \\
\hline
Customer (SF100) & 723751 & 4441.5 & 1220.0 & 11.1 \\
Part (SF100)     & 771981  & 2796   & 785.0   & 11.1 \\
Supplier (SF100) & 34582.25 & 690 & 67.0   & 11.1 \\
\hline
\end{tabular}
\end{table}
}

\emph{Skewed Join and Duplication Linked List effect:} JSPIM is particularly effective for scenarios involving right tables with duplication such as \texttt{lineorder} self-joins, as it remains unaffected by common challenges such as data skew in hash table and large table bucket sizes. The system achieves its most substantial improvements in joins with higher replication rates or data skew on both sides of the joins. Figure \ref{selfjoin} shows that, compared to DuckDB, JSPIM improves the execution time of joins on \texttt{lineorder} table columns 0 and 3 by up to 24241x. For these measurements, we configured DuckDB with a 12TB NVMe backup directory to prevent system memory constraints from terminating the process. However, due to resource limitations, we could only compute results for SF1 and SF10 in columns 0 and 3 on DuckDB, which have the lowest number of replicas, containing  $\sim$200K distinct entries.
\begin{figure}
    \centering
    \includegraphics[
    width=0.45\linewidth]{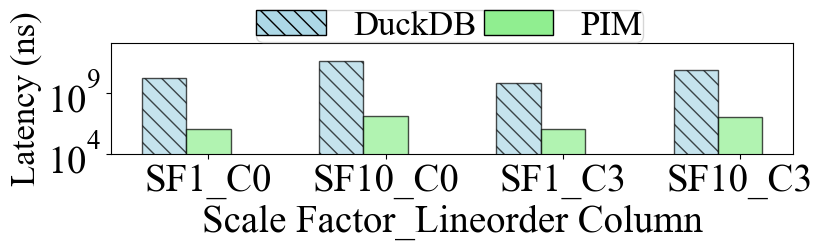}
    \caption{ JSPIM over DuckDB Self join latency comparison. {\footnotesize(\texttt{SELECT n.*, r.* FROM lineorder n JOIN lineorder r USING (column);})}. The columns have replicated words. (Y-axis in log scale)}
    \label{selfjoin}
\end{figure}

\emph{Select query results:} Figure \ref{fig:selectlatency} shows JSPIM's substantial performance advantages over DuckDB executing \texttt{select distinct} and \texttt{select where} queries. We performed the \texttt{select distinct} operations on shared columns between tables and computed the average performance. JSPIM's innovative duplication list mechanism enables it to achieve remarkable speedups—up to 403,300× faster than DuckDB when processing large tables (scale factor 100). Our evaluation of \texttt{select where} queries focused on equality conditions across diverse columns, with results averaged across multiple test cases. Since these queries require only a single search operation in JSPIM, their execution time remains constant, corresponding to one PIM cycle plus PIM activation. Under these conditions, JSPIM operats up to 1,938,461× faster than DuckDB when handling large datasets.
\begin{figure}
    \centering
    \includegraphics[
    width=0.6\linewidth]{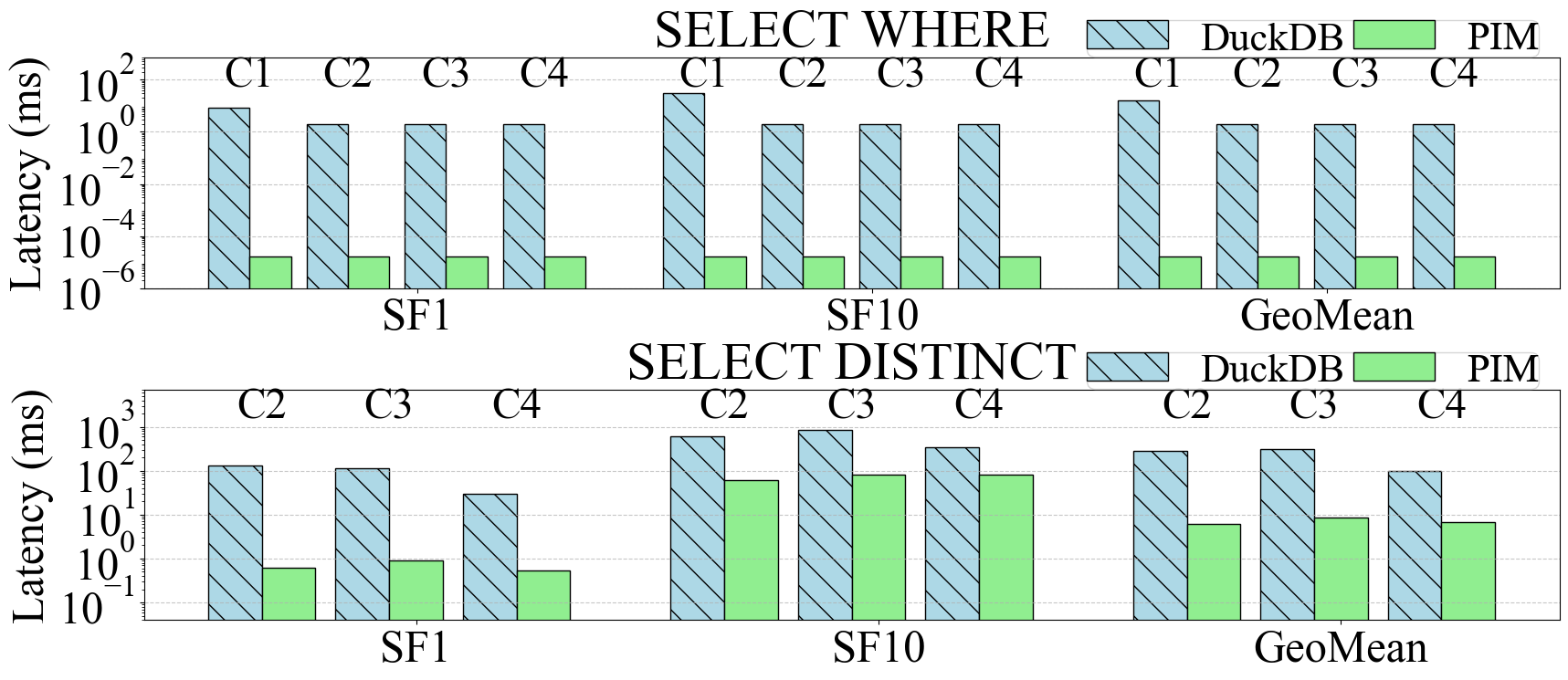}
    \caption{Comparison of JSPIM over DuckDB for queries: {\footnotesize (\texttt{SELECT j FROM tbl WHERE i = 3;}) and (\texttt{SELECT DISTINCT city FROM addresses;})} on Lineorder columns (c1-c4) - (Y-axis in log scale)}
    \label{fig:selectlatency}
\end{figure}

\subsubsection{Comparison with state-of-the-art PIM systems.}

Table~\ref{tab:tablecomppid} shows the comparison results of JSPIM with SPID, PID-Join across system design, hardware cost, and performance. PID-Join is sensitive to data skew, and both SPID and PID are limited by the WRAM storage size, leading to out-of-memory (OOM) failures in certain configurations. While no OOM occurs at 0.5M tuples, PID-Join fails at 8M tuples when Zipf $\geq$ 1.5. SPID-Join encounters OOM at 32M and 64M tuples (R and S) with Zipf = 2. We run JSPIM across same conditions to report its minimum and maximum speedups. All keys and values are 32 bits, with no compression on JSPIM.

JSPIM’s resilience to data skew and independence from hash bucket sizing enable significant speedups as Zipf skew and build table size grow. Even under low skew and small inputs, JSPIM achieves up to 15× speedup over SPID, due to reduced communication overhead, elimination of layout tuning, and its ability to stream results entry-by-entry—unlike SPID, which stalls until join completion.
For JSPIM, repeated keys further reduce row activations and lookup latency, amplifying performance gains.
\begin{table}
    \caption{Comparison of PID-Join, SPID-Join, and JSPIM.}
    \label{tab:tablecomppid}
    \centering
    \footnotesize
    \setlength{\tabcolsep}{2pt} 
    \renewcommand{\arraystretch}{0.95} 
    \begin{tabular}{p{2.6cm}p{3.55cm}p{3.5cm}p{3.5cm}}
        \hline
        \textbf{Parameter} & \textbf{PID-Join\cite{PIMJOINDRAM}} & \textbf{SPID-Join\cite{SPID}} & \textbf{JSPIM (Ours)} \\
        \hline
        \rowcolor{gray!7} \multicolumn{4}{c}{\textbf{System Settings}} \\
         \hline
        DRAM Type & DDR4 UPMEM & DDR4 UPMEM & DDR4 LRDIMM \\
        Area (per rank) & 64×(88KB RAM+PU) & 64×(88KB RAM+PU)  & 1 RLU + 105k comparators \\
        Level & Bank & Bank & Subarray \& Rank \\
        Channel / Ranks & 4 / 16 & 4 / 16 & 4 / 16 \\
           \hline
        \rowcolor{gray!7} \multicolumn{4}{c}{\textbf{Latency Analysis}} \\
           \hline
        Data Sensitivity & High (skew) & Low (distribute \& copy) & Not sensitive \\
        Data Movement & WRAM transfers & CPU-mediated inter-rank & In-place processing\\
        Layout Tuning & No & Required & No \\
           \hline
        \rowcolor{gray!7} \multicolumn{4}{c}{\textbf{Latency Speedup} (Zipf [0 , 2], $R\Join S \text{, } |s|\in [|R| \text{ , } 8\times|R|]$  )}\\
           \hline
        |R|= 32M & Baseline & [1,3]x PID & [20,300]x SPID \\
        |R|= 8M  & Baseline & [1,6]x PID& [15,100]x SPID\\
        |R|= 0.5M & Baseline & [1,10]x PID& [100,200]x SPID\\
        \hline
    \end{tabular}
\end{table}

\subsubsection{Comparison with GPU-based systems.}\label{4.2.3} 
As highlighted in Section~\ref{sec-join-algorithms}, while GPUs are often faster for hash joins, their advantages are conditional on specific workloads, while CPU-focused evaluation ensures applicability to broad scenarios. 
Figure \ref{GPU} compares the performance of JSPIM, GPU, and CPU. The results indicate that while JSPIM outperforms both the CPU and GPU, the GPU can become slower than the CPU as the data size grows, and spilling to main memory is inevitable. 
GPU join latency measurements exclude the query optimization setup phase ($\sim$0.03 s) and CPU–GPU data transfer times to isolate the record matching execution time.
\begin{figure}
    \centering
    \begin{subfigure}{0.8\linewidth}
        \centering
        \includegraphics[width=\linewidth]{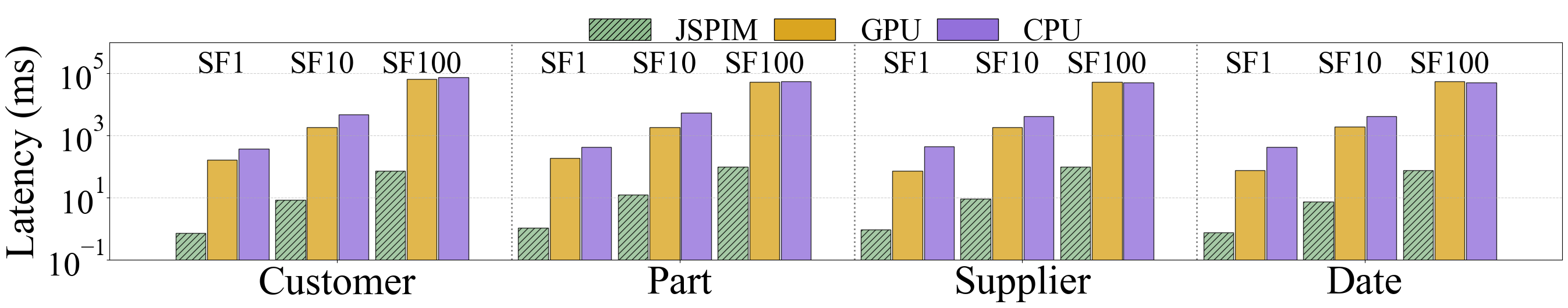}
    \end{subfigure}
    \hfill
    \begin{subfigure}{0.8\linewidth}
        \centering
        \includegraphics[ width=\linewidth]{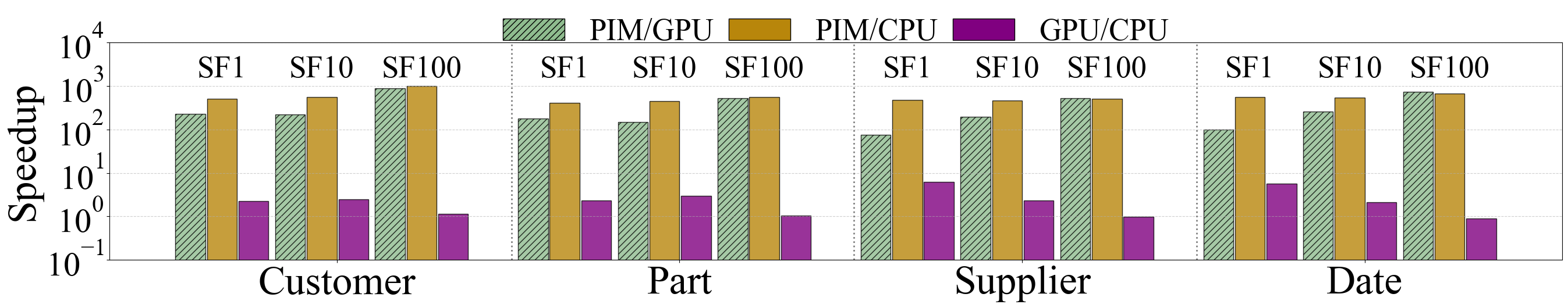}
    \end{subfigure}
    \caption{Latency and Speedup of JSPIM vs. GPU/CPU for Join Query. Measured on lineorder and dimension tables with an empty duplication list. GPU retains results; JSPIM sends key-value pairs to CPU. (Y-axis in log scale.)}
    \label{GPU}
\end{figure}
\subsubsection{Complete query optimization.}
To evaluate the system-level impact of integrating JSPIM, we compare full SSB query execution time in two modes: a baseline DuckDB execution (with all optimizations enabled), and a JSPIM-integrated mode where joins are offloaded to PIM. 
We model the integration of JSPIM by identifying the exact point in the DuckDB query plan where the join is invoked and replacing that portion with JSPIM’s 
join latency. We preserve the cardinality and filtering time reported by DuckDB’s profiling tools to ensure that the input/output sizes match realistic scenarios. Specifically, we use the number of rows flowing into the join to parameterize the JSPIM simulation and to estimate the volume of data transferred between PIM and CPU.

For example, in Query 1.1, both input tables are scanned and filtered on CPU, while the join is offloaded to JSPIM. In DuckDB’s native execution, the filter on the right table is applied before the join begins. In contrast, with JSPIM, this filter is applied on-the-fly during result streaming from PIM to CPU: each result index is checked against the filtered right-hand set as the data arrives. Despite this shift in when the filtering occurs, the impact on overall latency is minimal. We observe that the non-join portion of the query remains nearly unchanged over SSB queries, indicating that this overlap between filtering and streaming introduces negligible overhead.

Each query is executed under both Cold and Warm scenarios. In Cold runs, the query is executed without prior cache warming, while in Warm runs, we pre-execute the same query to allow caching effects to take place. For DuckDB, the difference between Cold and Warm is significant: Warm runs benefit from data structure reuse (e.g., hash tables, cached buffers). 
In contrast, JSPIM exhibits consistent join latency across both set of runs since the join execution is handled in a stateless fashion by PIM and does not reuse CPU-side caches or data structures. Hence, any Warm-vs-Cold speedup discrepancy originates from the non-join parts of the query.

Figure~\ref{fig:ssbquery} shows that integrating JSPIM yields up to 28× speedup over the baseline. As expected, the benefit is minimal for Q1-type queries, which perform negligible join work (Figure~\ref{fig:ssbquery}). In contrast, Q3.4 achieves the highest speedup due to two factors: it involves a large and costly join in the baseline, and the left table cardinality after filtering is significantly reduced, which minimizes data transfer from PIM to CPU. 
Some Q2 and Q4 queries, however, exhibit smaller speedups despite having $≥95\%$ of their execution time attributed to joins. This behavior stems from two main factors. First, in many of these queries, the left table’s cardinality remains high even after filtering, leading to more data transfer overhead from PIM to CPU. Second, the characteristics of the dimension tables involved in the joins differs. For example, in Q3.4, the join involves the Date table—a small dimension that fits entirely within a single PIM bucket. This causes many fact table entries to map to the same join key, resulting in high redundancy during processing. This allows sequential access patterns and low PIM-side latency. In contrast, Q2 queries involve joins with the larger Part table, which has less redundancy and a more scattered layout in PIM, resulting in higher access latencies.%
\begin{figure}
    \centering
    \includegraphics[
    width=0.8\linewidth]{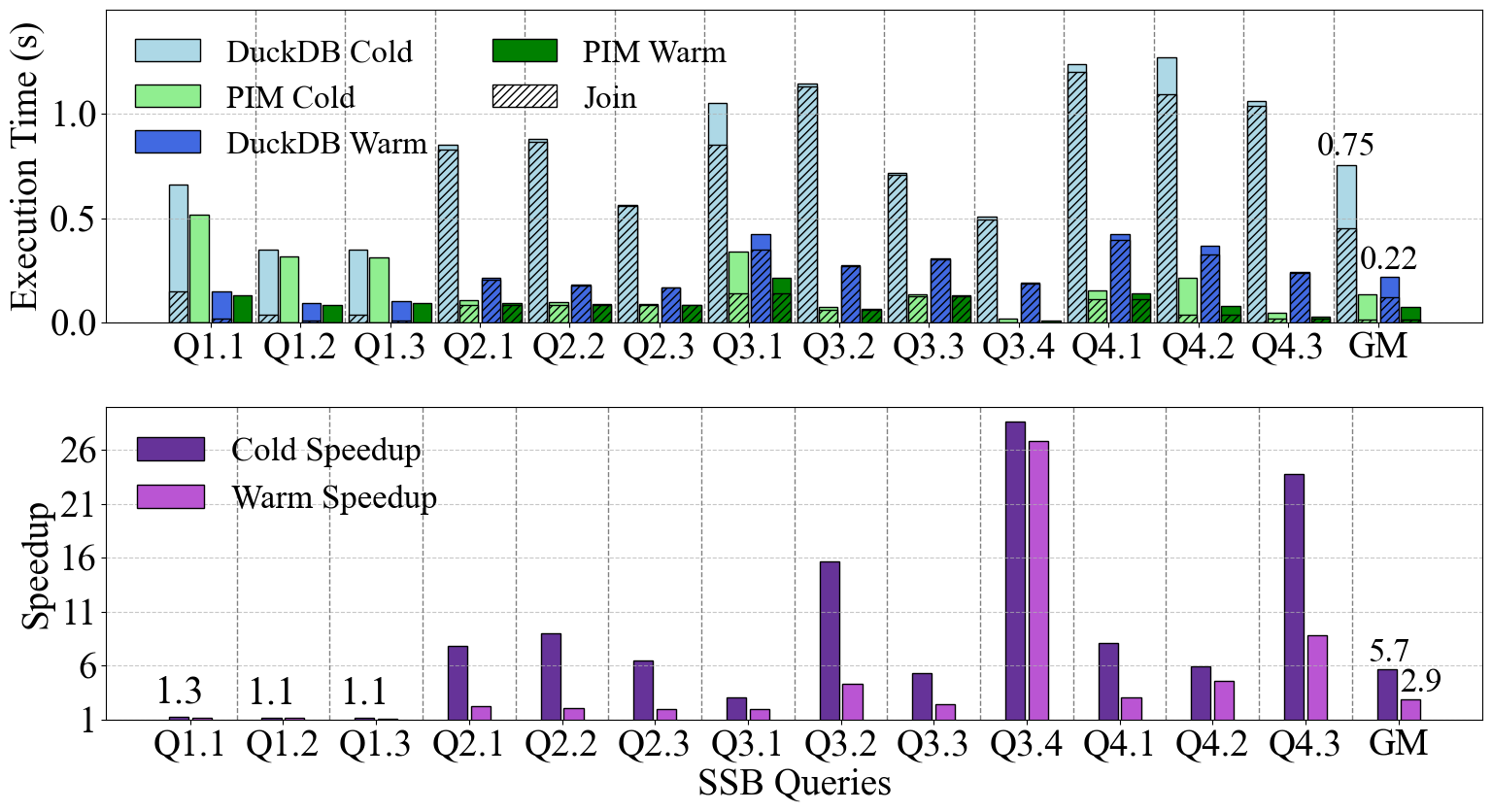}
    \caption{SSB query latency measurement of JSPIM integration with DuckDB compared to DuckDB using CPU for join computations. The hatched sections indicate join latency.}
    \label{fig:ssbquery}
\end{figure}

\subsection{Overhead and costs}
To evaluate the overhead of our design, we focus on two key factors: data overhead and area overhead (hardware cost).
\subsubsection{Data Overhead}\label{4.2.1}
JSPIM adds the following data structures: a dictionary, copies of encoded fact table key columns, a hash table, and a duplication-linked list.  
When the hash table is small, it can be replicated across DIMMs to reduce setup time. For larger datasets, the data is distributed across ranks, with separate hash tables generated for each dimension table. The fact table is partitioned to align with the relevant hash table segments.

To support select queries, relevant columns must be included in the hash and duplication tables, either as standalone tables or integrated into the dataset. Each supported column introduces additional data overhead.
In this design, only the hash table and fact table data are stored in PIM, while the remaining components are kept on the CPU.
For the evaluated SSB benchmark, the data overhead is 79.028 MB $\times$ SF in our experiments, which is about 7\% of the dataset size. 

\subsubsection{Area Overhead} \label{4.2.2}
JSPIM adds two key components to the DRAM module: the Rank-Level Unit (RLU) and subarray-level search units. To estimate the area overhead of our design, we compare it to the chip area of a simple MIPS processor described in~\cite{minimips}. The MIPS chip measures $30.8 , \mu m^2$ using 14nm technology, which is about $8.81 \times 10^{-7}$ times smaller than a DRAM chip with a size of $34 , mm^2$~\cite{hifidram}.
We implemented the subarray-level comparators and match-select units in RTL and used the Synopsys Design Compiler with 14nm technology to evaluate their power and area. The subarray architecture consists of multiple comparators connected to row buffers and a match-select unit. Using the Synopsys tool~\cite{synopsys,synopsys_cloud_2024}, we found that the subarray-level design has an area overhead of $726835.2 \mu m^2$ per chip in 14nm technology. This is about 50 times smaller than a DRAM chip and adds only 2.13\% area overhead to the chip.
{While 14nm transistors may be smaller than those used in DRAM (as observed in HiFi-DRAM~\cite{hifidram}), the real transistor size and area overhead in DRAM could be up to 1.5x or 2x larger due to layout constraints, such as the need for additional wiring, spacing for reliability, and the shared nature of the sense amplifier and subarray regions~\cite{hifidram}.}

Adding the subarray module increases power consumption by $0.584 , \mu W$ per active subarray. Even assuming one active subarray throughout the join computation and considering that the RLU consumes as much power as a CPU in the worst case, our design still achieves significant energy savings. This is because our latency improvement exceeds 100x compared to the basic CPU method.

\subsubsection{Sensitivity Analysis}
As reported in prior work~\cite{sive} and confirmed by Synopsys, the \textbf{subarray-level architecture }adds less than one cycle of latency without affecting standard DRAM timing.  However, to assess its impact, we model a fixed per-read delay 
$t_{\text{CMP}}$ in DRAMSim3, sweeping it from 0 to 4 cycles while measuring join latency across SSB scale factors. 
 Once the subarray delay exceeds the burst cycle, it becomes a dominant constraint~\cite{utahslides} in the controller’s scheduling decisions, altering the order of requests and the timing of activate/precharge commands. 
 Increasing $t_{\text{CMP}}$ from 0 to 1 cycle increases the join latency by 11\% (averaged across SSB tables pure join, for SF1, SF10, SF100). 
 Additional delays beyond this point have diminishing impact, as the pipeline is already stalled, with an average join latency increase of 32\% at $t_{\text{CMP}}=4$
and a maximum of 61\% for the customer-lineorder join at SF1. 
This reduces JSPIM’s pure join query speedup over DuckDB from 400×–1000× to 324×–916×. When integrated with DuckDB to run full SSB SF100 queries, the speedup range changes from 1.1×–28× to 1.1×–27×. 
The results are shown in Figure~\ref{sens}.
\begin{figure}
    \centering
    \includegraphics[width=0.5\linewidth]{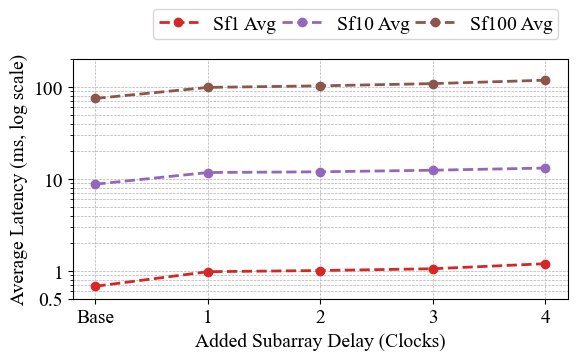}
    \caption{Sensitivity of average join latency to subarray-level delay over SSB scale factors for $(Lineorder \Join Dimention)$.}
    \label{sens}
\end{figure}

We analyze \textbf{sensitivity to the number of ranks} with 32 DIMMs, comparing one versus two ranks per DIMM. Performance improves with added ranks due to intra-DIMM parallelism, but gains are sublinear as ranks share bandwidth, limiting CPU data transfer.

\section{Conclusion}
We designed and evaluated JSPIM, a platform that leverages Load-Reduced DIMM (LRDIMM) PIM architectures to optimize join queries through both architectural and algorithmic innovations. 
JSPIM provides constant-time search capabilities by using acceleration at the subarray level to exploit 
parallelism, eliminates costly cross-chip communication by using rank-level pipelined processing, and 
overcomes the data skew problem by synergistic algorithm-hardware data design. 
JSPIM integration with DuckDB demonstrates 
{a range of 1.1x to 28x} speedup in end-to-end query performance compared to DuckDB running on a CPU, with a 7\% increase in data and 2.1\% area overhead. 
Our results demonstrate the potential for PIM to mitigate the impact of the memory wall on the performance of database queries.

\section{Acknowledgement}
This work was supported in part by NSF grant award CNS-1900803.

\bibliographystyle{ieeetr}
\bibliography{refs}

\end{document}